%% file: main.tex
\documentclass[10pt,twocolumn,letterpaper]{article}

\usepackage[pagenumbers]{iccv} % To force page numbers, e.g. for an arXiv version


\definecolor{iccvblue}{rgb}{0.21,0.49,0.74}
\usepackage[pagebackref,breaklinks,colorlinks,allcolors=iccvblue]{hyperref}
\usepackage{makecell}
\usepackage{multirow}
\usepackage{algorithm}
\usepackage{algpseudocode}
\usepackage{amsmath}
\usepackage{booktabs}
\usepackage{amsfonts}
\usepackage{multicol}
\newcommand{\tensor}[1]{\mathbf{#1}}
\newcommand{\set}[1]{\mathcal{#1}}
\newcommand{\field}[1]{\mathbb{#1}}
\newcommand{\op}[1]{\text{#1}}
\def\paperID{xxx} % *** Enter the Paper ID here
\def\confName{CVPR}
\def\confYear{2026}

\title{ReLumix: Extending Image Relighting to Video via Video Diffusion Models}

\author{Lezhong Wang\textsuperscript{1} \quad Shutong Jin\textsuperscript{2} \quad Ruiqi Cui\textsuperscript{1} \quad \\ \quad Anders Bjorholm Dahl \textsuperscript{1} \quad Jeppe Revall Frisvad\textsuperscript{1} \quad Siavash Bigdeli\textsuperscript{1}\\ 
\textsuperscript{1}Technical University of Denmark \quad \textsuperscript{2}KTH Royal Institute of Technology \\ 
{\tt\small \{lewa, ruicu, abda, jerf, sarbi\}@dtu.dk, shutong@kth.se}}

\begin{document}
\twocolumn[{%
\renewcommand\twocolumn[1][]{#1}%
\maketitle
\begin{center}
    \vspace{-2ex}
    \centering
    \includegraphics[width=1.0\textwidth]{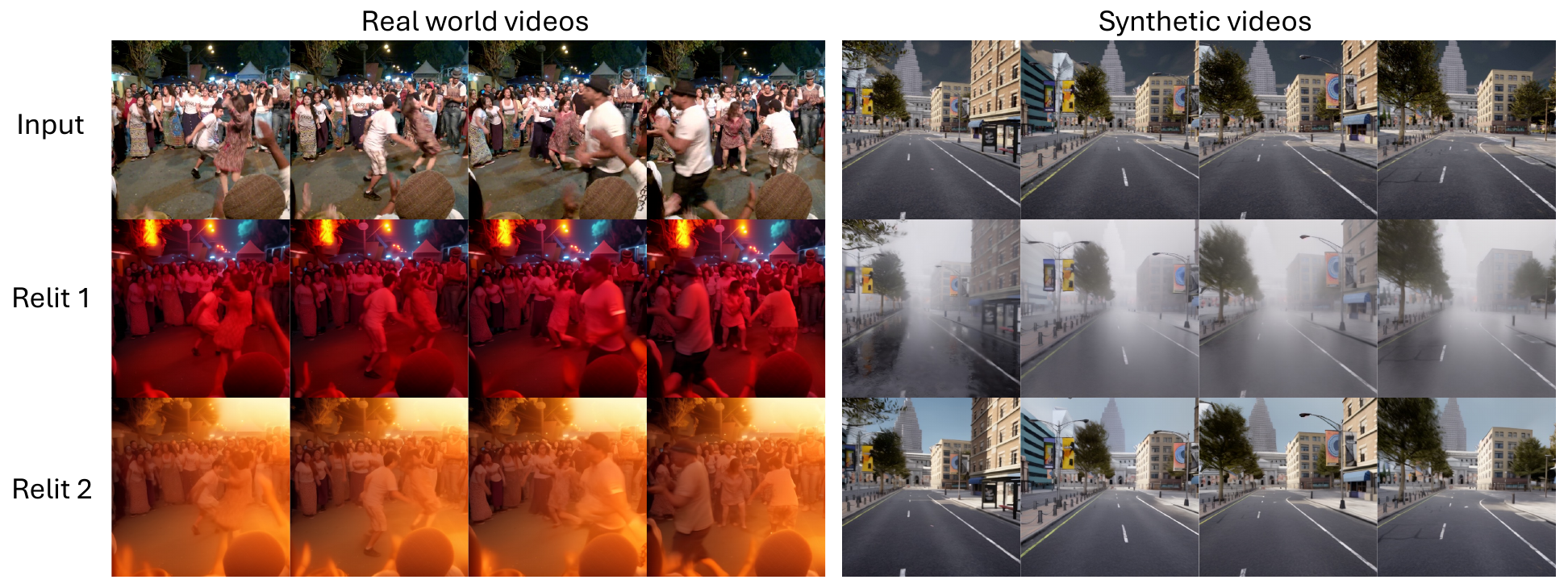}  \\[-2ex]
    \captionof{figure}{The user selects one video frame as the reference and changes this using any preferred relighting software or method. Relumix then propagates these edits across the remaining frames to complete the video editing process. Our approach is trained on the CARLA synthetic dataset, enabling the model to learn the intrinsic representation of light and shadow even from basic data. This results in strong zero-shot capabilities, allowing the model to be applied to real-world videos without any fine-tuning.
    }
    \label{fig:teaser}
\end{center} }]

\input{sec/0_abstract}    
\input{sec/1_intro}
\input{sec/2_body}

{
    \small
    \bibliographystyle{ieeenat_fullname}
    \bibliography{main}
}
\clearpage
\input{sec/X_Appendix}

\end{document}

%% file: sec/0_abstract.tex
\begin{abstract}
Controlling illumination during video post-production is a crucial yet elusive goal in computational photography. Existing methods often lack flexibility, restricting users to certain relighting models. This paper introduces \textbf{ReLumix}, a novel framework that decouples the relighting algorithm from temporal synthesis, thereby enabling any image relighting technique to be seamlessly applied to video. Our approach reformulates video relighting into a simple yet effective two-stage process: (1) an artist relights a single reference frame using any preferred image-based technique (e.g., Diffusion Models, physics-based renderers); and (2) a fine-tuned stable video diffusion (SVD) model seamlessly propagates this target illumination throughout the sequence. To ensure temporal coherence and prevent artifacts, we introduce a gated cross-attention mechanism for smooth feature blending and a temporal bootstrapping strategy that harnesses SVD's powerful motion priors. Although trained on synthetic data, ReLumix shows competitive generalization to real-world videos. The method demonstrates significant improvements in visual fidelity, offering a scalable and versatile solution for dynamic lighting control.
\end{abstract}

%% file: sec/1_intro.tex
\section{Introduction}
\label{sec:introduction}
In the realm of visual perception, illumination conditions serve as a foundational element that governs both the aesthetic and interpretative dimensions of scenes. This critical role has driven extensive research into illumination editing techniques, which remain pivotal in fields ranging from entertainment to virtual environment design. Recent advancements in generative models~\citep{rombach2022high,guo2023animatediff,blattmann2023stable,wang2023modelscope} have demonstrated potential in automated video processing, offering substantial reductions in labor-intensive manual workflows. However, the challenge of dynamically manipulating lighting conditions in video sequences remains a complex problem due to the interplay of temporal coherence and spatial consistency requirements.

Two primary paradigms have emerged in image-based illumination editing research. The first involves data-centric methodologies~\citep{deng2025flashtex, ren2024relightful, kim2024switchlight, zhang2024lumisculpt, zeng2024dilightnet, jin2024neural}, which combine extensive relighting datasets with pre-trained diffusion architectures. These approaches, exemplified by IC-Light~\citep{iclight}, leverage light transport independence to preserve surface reflectance while modifying illumination. The second paradigm employs physics-based illumination models \citep{wu2023factorized,yao2022neilf,wang2025materialist}, which simulate light-material interactions through physical renderers. While these methods offer superior interpretability and precision, their computational optimization of single-image demands often limits practical applications in videos.

Moreover, the transition from image-based to video-based relighting presents non-trivial challenges, primarily due to significant computational bottlenecks. For instance, methods like Light-A-Video~\cite{zhou2025light} require frame-by-frame inversion processes, resulting in prohibitively long inference times. Additionally, neural rendering approaches demand substantial computational resources for training, as demonstrated by DiffusionRenderer~\cite{liang2025diffusionrenderer} which utilizes 150,000 annotated frames and 32 A100 GPUs for two days of training. Meanwhile, physics-based methods remain constrained by the high cost of differentiable rendering, with single-frame material property estimation requiring minutes of optimization, thus unsuitable for direct usage in video-based relighting.

This study introduces a novel framework that bridges these paradigms, enabling efficient video relighting through a modular architecture. Our approach decouples the relighting process into two stages: (1) applying any image-based relighting technique to a reference frame, and (2) propagating the lighting changes across the video sequence using a fine-tuned stable video diffusion (SVD) model. This design enables seamless integration of both data-driven and physics-based methods while addressing key limitations in speed and data efficiency.

The primary contributions of this work are:
\begin{itemize}
    \item \textbf{Modular and Flexible Framework}: By decoupling reference frame relighting from temporal propagation, our method supports arbitrary image relighting techniques, including both neural rendering approaches like IC-Light and physics-based methods such as Materialist~\cite{wang2025materialist}.
    \item \textbf{Enhanced Video Generation Efficiency}: Our approach eliminates the need for frame inversion, achieving 9× speed-up over I2VEdit~\cite{ouyang2024i2vedit} and 6× over Light-A-Video~\cite{zhou2025light} while maintaining great temporal consistency and visual quality.
    \item \textbf{Effective Sim-to-Real Generalization from Synthetic Data}: We demonstrate that fine-tuning for only 12 hours on a single H100 using a synthetic-only dataset, CARLA \cite{dosovitskiy2017carla}, is sufficient to achieve robust, high-fidelity relighting on diverse real-world videos. This approach bypasses the need for expensive and hard-to-acquire large-scale annotated real-world video data, showcasing strong sim-to-real transfer capabilities.
    \item \textbf{Novel Mechanisms for Mask-Free Propagation}: We introduce architectural innovations that achieve high-fidelity propagation by implicitly learning to separate content from illumination. This eliminates the need for explicit segmentation masks common in classic relighting pipelines \cite{guo2025high}, greatly enhancing practical usability. This mask-free design robustly prevents artifacts like identity drift and flicker, even in scenes with complex object interactions.
\end{itemize}

Experiments demonstrate that ReLumix achieves state-of-the-art performance in video relighting, surpassing existing methods in consistency, quality, and adaptability, see examples in Fig.~\ref{fig:teaser}. Unlike most other methods \cite{zhou2025light,guo2025high}, which is constrained by a fixed relighting model, ReLumix allows seamless integration of any relighting technique, making it highly future-proof.

%% file: sec/2_body.tex
\section{Related Work}
\label{sec:related}

\subsection{Video Generation and Editing with Diffusion Models}
Video diffusion models have rapidly advanced, evolving from text-to-video (T2V) synthesis to more controllable generation and editing tasks. Early T2V models successfully extended text-to-image architectures with temporal modules to ensure frame consistency, as exemplified by works like AnimateDiff~\cite{guo2023animatediff}. A significant development has been the rise of image-to-video (I2V) models, such as Stable Video Diffusion (SVD) ~\cite{blattmann2023stable}, which excel at animating still images with realistic motion by fine-tuning on large-scale video data.
This I2V paradigm has become a cornerstone for video editing. Many state-of-the-art methods operate by propagating edits from a single user-modified frame across the entire sequence. For instance, approaches like TokenFlow~\cite{geyer2023tokenflow} and MagicProp~\cite{yan2023magicprop} achieve this through sophisticated attention mechanism manipulation or feature injection, ensuring that the original video's motion and content are preserved. I2VEdit \cite{ouyang2024i2vedit} also adopts an image-to-video editing approach, though its method is highly complex. It employs attention matching and inverted latents to ensure motion and appearance consistency, which requires per-video adjustments and facing challenges in generalization.
While these general ``edit-propagate" methods, including recent works like I2VEdit \cite{ouyang2024i2vedit} and LoRA-Edit \cite{gao2025lora}, provide a powerful foundation, they are not inherently optimized for the nuances of relighting. Illumination is a global property that interacts with scene geometry and materials in complex ways. Our work specializes this paradigm, introducing specific mechanisms to ensure that only the lighting, not the underlying content or motion is altered, a challenge that remains for general-purpose editors.

\subsection{Learning-based Illumination Editing}
Learning-based methods now dominate the field of illumination control, far surpassing traditional models that struggle with real-world material and lighting estimation. Research has produced powerful models, such as Diffusion Model \cite{he2024diffrelight,mei2025lux,wang2025comprehensive,fang2025relightvid,he2025unirelight} for relighting various subjects, with a notable focus on general images. For example, recent techniques like IC-Light~\cite{iclight} demonstrate the ability to disentangle and modify illumination from a single image while preserving object albedo with remarkable fidelity. In the video domain, controlling lighting is significantly more challenging due to the temporal consistency constraint. While some methods ~\cite{zhang2024lumisculpt,liu2025tc} have explored lighting control within T2V generation. Recently, numerous concurrent works \cite{liang2025diffusionrenderer,he2025unirelight} have sought to solve the problem and start to use environment maps for more precise lighting control.
However, these methods are often tied to a specific model architecture. There is a clear gap for a method that can leverage the power of state-of-the-art image relighting models and apply them consistently to video. Our framework directly addresses this by integrating such image-based techniques into a robust video propagation pipeline.

\subsection{Physically-based Illumination Editing}
Physically-based inverse rendering methods offer strong interpretability and realism, enabling precise control over illumination effects such as shadows, reflections, and refractions. However, these advantages have traditionally come at the cost of input requirements and computational overhead. Early approaches and even many modern techniques rely on multi-view inputs to robustly solve the ill-posed problem of decomposing a scene into its intrinsic components~\cite{azinovic2019inverse}. The advent of Neural Radiance Fields (NeRF)~\cite{mildenhall2021nerf} and differentiable rendering~\cite{luan2021unified, zhang2022iron} has made multi-view decomposition more feasible, yielding impressive results in material and lighting estimation~\cite{li2023multi, sun2023neural}. 
A common thread among these methods is a time-consuming, per-scene optimization phase to estimate material properties before relighting can occur. While recent work like Materialist~\cite{wang2025materialist} have extended this paradigm to the more challenging single-view setting, they still require a significant optimization process for each new image. This computational demand hinders their application in interactive or large-scale video processing tasks. In contrast, our framework is designed to bypass this costly per-frame optimization, offering a flexible and efficient alternative for video relighting.

\subsection{Image Editing Foundations}
Our work is fundamentally enabled by the tremendous progress in diffusion-based image editing. A rich ecosystem of tools now exists to modify images with high precision \cite{shi2024dragdiffusion,hertz2022prompt,kawar2023imagic,wang2024stereodiffusion,pan2023drag}. These range from zero-shot, prompt-based methods like Prompt-to-Prompt~\cite{hertz2023prompt} to personalization techniques like DreamBooth~\cite{ruiz2023dreambooth} and instruction-guided editors like Instruct-Pix2Pix~\cite{brooks2023instructpix2pix}. The maturity and diversity of these image editing methods provide a strong foundation for our two-stage approach. By leveraging these powerful tools for the reference frame relighting step, we can focus our core contribution on the unsolved challenge of temporally consistent propagation.

\begin{figure}[t]
    \centering
    \includegraphics[width=0.9\linewidth]{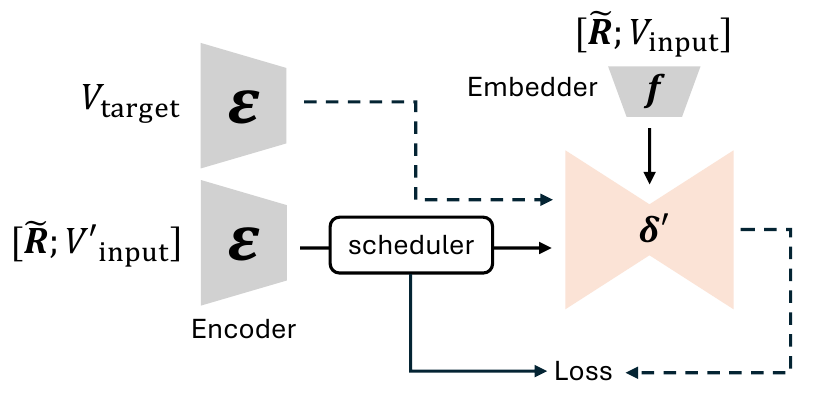}
    \caption{Our proposed training method. Here, \( V'_{\text{input}} \) represents the input video \( V_{\text{input}} \) with the reference frame \(\mathbf{R} \) replaced, \([;]\) denotes the concatenation operation, \(\mathbf{\tilde{R}}\) represents the replicated reference frame \(\mathbf{\tilde{R}} \in \mathbb{R}^{T\times H\times W\times 3}\), and \(\delta'\) denotes the naive Cross Attention of Denoising UNet replaced by the Gated Cross Attention module (Sec.~\ref{sec:GCA}). Note that the figure only includes the components we modified. The modules that SVD originally had, such as \textit{TimestepEmbedder}, are not shown.}
    \label{fig:method}
\end{figure}

\section{Method} \label{sec:method}

Our framework enables video relighting by introducing a novel adaptation strategy for SVD, leveraging synthetic data while preserving strong generalization to real-world scenarios.
During training, models observe a source video $\boldsymbol{V}_{\mathrm{input}}$ from one camera view under random initial lighting $l_{\mathrm{src}}$ and must generate the same scene from the \textit{same} camera view under target lighting $l_{\mathrm{target}}$ (see Fig.~\ref{fig:method}). 

Our key focus is to ensure that the model learns an intrinsic representation of lighting, specifically how light interacts with objects rather than merely overfitting to a specific dataset. The goal is to extend this learned understanding to entirely unseen domains without requiring additional fine-tuning.
Our pipeline consists of three key innovations: conditional embedding fusion, gated cross-attention mechanisms, and temporal consistency enforcement.

\begin{figure}[!tb]
    \centering
    \includegraphics[width=\linewidth]{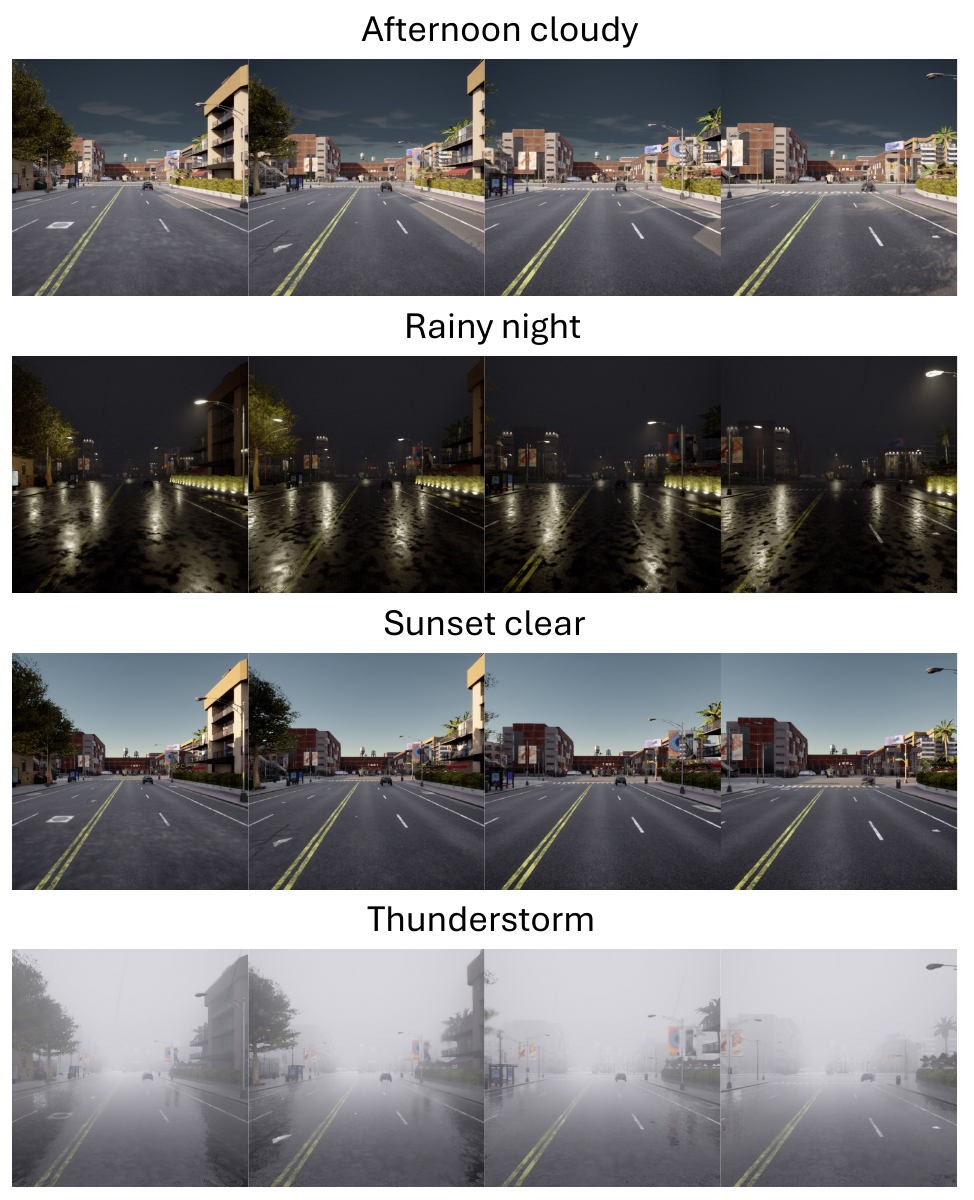}
    \caption{Examples of the synthetic dataset generated using the CARLA simulator.}
    \label{fig:dataset-examples}
\end{figure}

\subsection{CARLA Synthetic Dataset Preparation}   \label{sec:data}  

We choose the CARLA simulator ~\cite{dosovitskiy2017carla} as the tool for generating synthetic data due to its user-friendly nature, open-source structure, and widespread community support. CARLA is capable of generating complex, dynamic urban environments. Most importantly, it enables us to independently control the weather conditions while keeping other factors constant, which is highly aligned with the requirements of our method. We name our generated dataset \textit{CARLA Relight}. This dataset provides realistic driving scenarios with complex vehicle dynamics, unpredictable traffic flows, and diverse urban layouts.

Our data generation pipeline operates in two main stages to ensure consistency across varying conditions. First, for each scene, we record a 75 second vehicle trajectory. A primary vehicle, controlled by an intelligent agent, navigates through a town while interacting with 10 to 20 other non-player character (NPC) vehicles, which also operate under agent control to create natural traffic patterns. The complete set of transformations (location, rotation) and velocities for all vehicles is recorded at each timestep. Second, this recorded trajectory is replayed multiple times. In each replay, the vehicles follow the exact same paths, but we apply a different, randomly selected 10 weather and lighting configuration from a predefined pool of 15 conditions. These conditions include diverse times of day (e.g., noon clear, sunset cloudy, foggy night) and weather events (e.g., heavy rain, thunderstorm).

For each scene-lighting pair, we capture 1500 frames at evenly spaced intervals along the replayed trajectory. This \textit{record-and-replay} methodology guarantees that for a given scene, the underlying geometry, vehicle positions, and camera viewpoints are identical across all lighting variations, perfectly isolating illumination as the only variable. We generated data across 9 distinct CARLA towns, creating a total of 75 unique scenes (35 in the high-detail \textit{Town10HD} and 5 in each of the other 8 towns). This process yields over 1.12 million frames (75 scenes $\times$ 10 lighting variations $\times$ 1500 frames), providing a large-scale, highly realistic dataset for training and evaluating video relighting. 
We refer to  Fig.~\ref{fig:dataset-examples} and the supplementary video (Part 2) for examples.

\subsection{Temporal Bootstrapping via Frame Replacement}
To ground the video generation process in the artist's specific intent, we employ a simple yet highly effective bootstrapping strategy. The user-provided relit reference frame, $\tensor{R}$, physically replaces the first frame of the original input video, $\tensor{V}_{\mathrm{input}}$:
\begin{equation}
    \tensor{V}_{\mathrm{input}}' = [\tensor{R}, \tensor{V}_{\mathrm{input}}[1:T]]
    \label{eq:frame_replacement}
\end{equation}
where $T$ is the total number of frames. This modified sequence is then fed into the model.

This technique treats the reference frame as a strong \textbf{initial boundary condition}. In temporally-aware models like SVD, which generate frames autoregressively, the visual information from the first frame establishes a powerful prior that influences the synthesis of all subsequent frames \cite{blattmann2023stable}. By setting this initial condition to the artist's target, we anchor the entire generation process to the desired look, ensuring the intended illumination is propagated forward while preserving the original motion dynamics from the rest of the video. This strategy thus effectively propagates the target illumination throughout the video while preserving the original motion dynamics from $\tensor{V}_{\mathrm{input}}[1:T]$.

\subsection{Reference-Guided Embedding Fusion}
To ensure the fine-grained details of the target lighting---such as color palette, shadow softness, and texture---are faithfully reproduced, we introduce an early-stage embedding fusion. This provides the model with a constant, low-level visual guide throughout the denoising process.

First, we replicate the single reference frame $\tensor{R}$ across the time dimension to create a constant reference video $\tensor{\tilde{R}} \in \field{R}^{T\times H\times W\times 3}$. This tensor is then concatenated with the noised input video $\tensor{V}_{\mathrm{noisy}}$ along the channel axis. A pretrained video feature extractor $\set{E}$ processes this combined input:
\begin{equation}
    \tensor{E}_f = \set{E}([\tensor{\tilde{R}}, \tensor{V}_{\mathrm{noisy}}])
    \label{eq:embedding_fusion}
\end{equation}
The resulting fused embedding $\tensor{E}_f$ is then passed to the diffusion model's U-Net. This early-stage fusion acts as a strong conditioning signal, forcing the model's initial layers to recognize both the scene's content (from $\tensor{V}_{\mathrm{noisy}}$) and the target illumination's visual characteristics (from $\tensor{\tilde{R}}$). The persistent presence of the reference frame's features prevents the model from drifting away from the target appearance, especially in complex scenes or during long video sequences.

\subsection{Gated Cross-Attention for Temporal Consistency}
\label{sec:GCA}

\begin{algorithm}[!tb]
\caption{Gated Cross-Attention}
\label{alg:gated_attn}
\begin{algorithmic}[1]
\Require Context embedding $\tensor{C}$; Query Q from U-Net; Learnable static gate $\alpha$
\Ensure Gated output embedding $\tensor{O}$
\State $\tensor{O}_{\mathrm{attn}} \leftarrow \op{CrossAttention}(Q, \op{context}=\tensor{C})$ 
\State $\tensor{Q}_{\mathrm{global}} \leftarrow \op{mean}(Q, \op{dim}=1)$ 
\State $\tensor{G}_{\mathrm{dyn}} \leftarrow \sigma(W \cdot \tensor{Q}_{\mathrm{global}} + b)$ \Comment{Compute dynamic gate from query}
\State $\tensor{G} \leftarrow \alpha \cdot \tensor{G}_{\mathrm{dyn}}$ \Comment{Combine with learnable static gate}
\State $\tensor{O} \leftarrow \tensor{G} \odot \tensor{O}_{\mathrm{attn}}$ 
\State \textbf{return} $\tensor{O}$
\end{algorithmic}
\end{algorithm}

A key challenge in relighting is that the influence of the new illumination should not be uniform. For instance, a bright neon light should strongly affect diffuse surfaces but might only appear as a small specular highlight on a glossy material. To give the model this nuanced, content aware control, we introduce a Gated Cross-Attention (GC) mechanism. This module acts as an intelligent switch, adaptively modulating how much of the reference lighting style is applied at different locations and times.

As detailed in Algorithm~\ref{alg:gated_attn}, the query Q, which represents the spatial features from the diffusion model's U-Net, attends to the context $\tensor{C}$ to produce an attention output $\tensor{O}_{\mathrm{attn}}$. However, instead of using this output directly, we introduce a gating mechanism to scale its contribution. A dynamic gate value, $\tensor{G}_{\mathrm{dyn}}$, is first computed from the global features of the query Q itself via a small neural network.
This allows the model to learn to scale the influence of the cross-attention output based on the semantic content it is currently generating. 
This dynamic value is then multiplied by a learnable scalar parameter $\alpha$ to form the final gate $\tensor{G}$. This gate adaptively modulates the attention output to produce the final fused output $\tensor{O}$.

This architecture enables the model to dynamically decide how much to rely on the conditioning context, effectively learning an implicit mask for applying the relighting effect, which is critical for preventing artifacts and preserving the underlying identity of objects in the scene.

\begin{table}[!tb]
    \centering
    \caption{Quantitative evaluation and ablation study over \textit{CARLA Relight}. I2VEdit and our method use the GT's first frame as input, while DiffusionRenderer employs an envmap to relight the video. Therefore, the results of DiffusionRenderer are for reference only.}
    \begin{tabular}{c|ccc}
    \toprule
     Methods& SSIM $\uparrow$& PSNR ↑& LPIPS ↓\\ 
     \midrule
     DiffusionRenderer \cite{liang2025diffusionrenderer} & 0.585 & 13.353 & 0.255 \\
     \hline
     \rule{0pt}{2ex}I2VEdit \cite{ouyang2024i2vedit} & 0.611 & 18.135 & 0.379 \\
    \hline
    Ours (w/ FR) & 0.522 & 14.035 & 0.612 \\
    Ours (w/ FR\&EF) & 0.524 & 14.380 & 0.629 \\
    Ours (w/ FR\&GC) & 0.501 & 10.658 & 0.508 \\
    Ours (w/ FR\&GC\&EF) & \textbf{0.747} & \textbf{22.308} & \textbf{0.143} \\
    \bottomrule
    \end{tabular}

    \label{tab:CARLA_compare}
\end{table}

\subsection{Finetuning Details}
We initialize our model with the weights from SVD~\cite{blattmann2023stable} (14-frame version). Since the video relighting task is very different from the original SVD task, we do not use LoRA for fine-tuning, but unfreeze all the weights except for the embedders. The model is trained on the CARLA dataset for 10,000 steps using one H100 GPU, with a total training time of approximately 12 hours. Training is conducted on 30-frame video clips at a resolution of $512 \times 512$, with a batch size of 1. TF32 precision is enabled to accelerate training speed. We employ the AdamW optimizer ($\beta_1=0.9$, $\beta_2=0.95$) with a learning rate of $10^{-5}$ and gradient clipping at 1.0 to ensure stable training dynamics. 

For inference, we utilize the EDM sampler~\cite{karras2022elucidating} with 25 denoising steps and a classifier-free guidance weight of $w=1.5$. 
The combination of these techniques allows our model to achieve robust performance in relighting tasks while generalizing effectively to complex real-world scenarios.

 \begin{figure*}[!tbh]
     \centering
     \includegraphics[width=\linewidth]{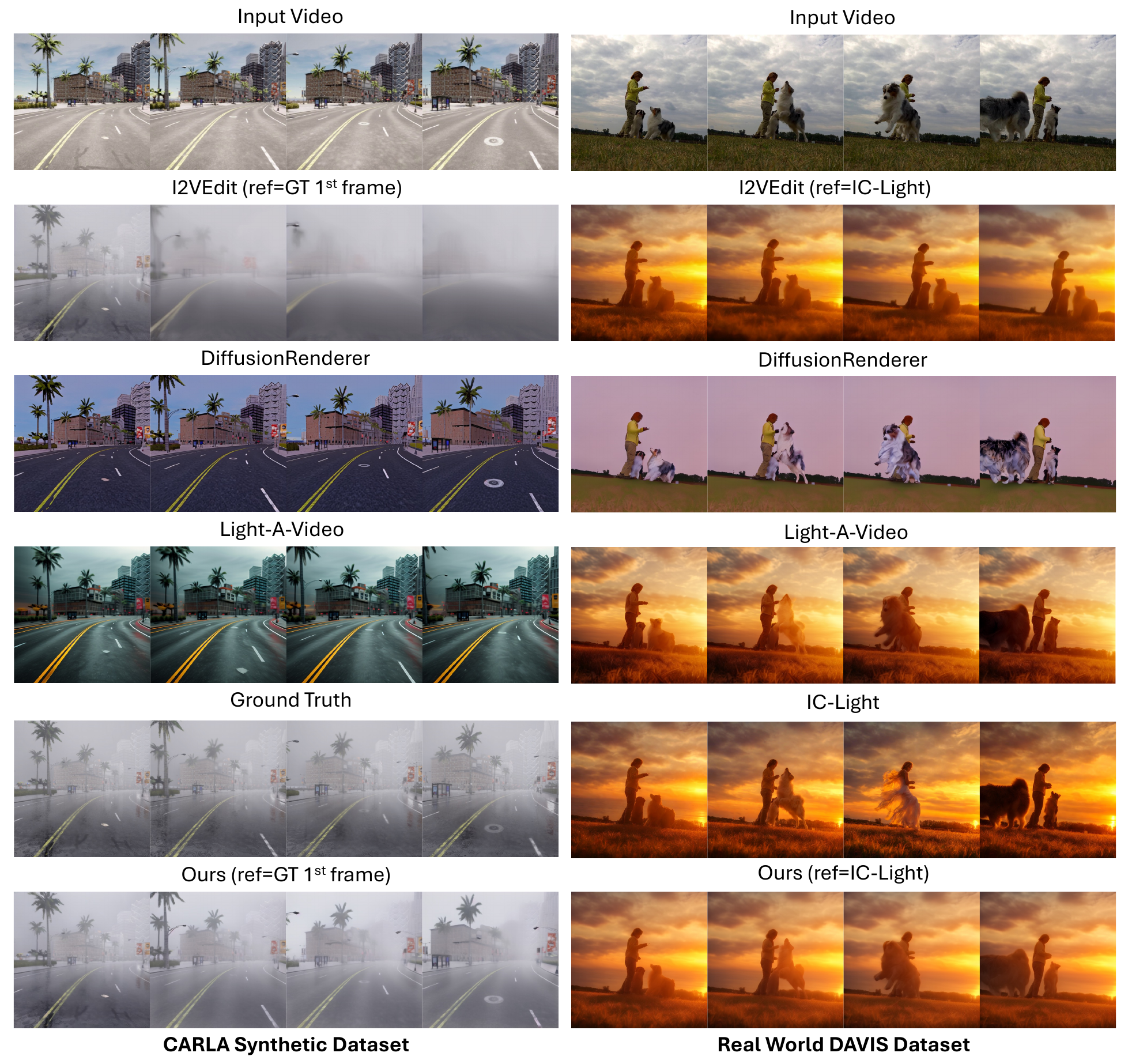}
     \caption{Qualitative evaluation of different methods on synthetic and real-world datasets. Note that these methods use different input types: DiffusionRenderer uses an envmap for relighting, while Light-A-Video and IC-Light use text prompts. I2VEdit and our method use a relit first frame for relighting. To demonstrate our method's strong generalizability to real-world conditions, we specifically selected videos for relighting that feature lighting effects considerably distinct from those in the \textit{CARLA Relight} dataset. This highlights our method's ability to perform well even with unseen lighting variations. }
     \label{fig:qual_compare}
 \end{figure*}
\section{Experiments} \label{sec:exp}  

\begin{table*}[!htb]
    \centering
    \caption{Quantitative evaluation and human evaluation results on real world videos (DAVIS). Comparison of different methods based on various evaluation metrics. Time represents the average processing time (in seconds) per frame on a single H100 (80GB) GPU. \textbf{Bold} number means the best, \underline{underscore} means the second best. Our method consistently achieves either the best or second-best results across various metrics in both quantitative and human evaluations.}
    \setlength{\tabcolsep}{1mm}
    \begin{tabular}{c|cccccc|c}
        \toprule
        \multirow{3}{*}{Method}&\multicolumn{6}{c|}{Quantitative Evaluation} & \multicolumn{1}{c}{Human Evaluation} \\
        \cline{2-8}
        & \makecell{\rule{0pt}{2ex}Temporal \\ LPIPS↓} & \makecell{Flow \\ Consistency↑} & \makecell{Color \\ Consistency↓} & \makecell{Brisque ↓}&\makecell{CD-FVD ↓} & Time (s)↓ & \makecell{\rule{0pt}{2ex}Video Relit \\ Satisfaction↑} \\
        \midrule
        Input (Reference) & 0.1781 & 0.1644 & 9.7437 & 24.5461 & - & - & 5  \\
        \hline
        IC-Light \cite{iclight} & 0.2948 & 0.0529 & 22.5074 & 36.6084 & 808.4 & 2.9 & 2.34  \\
        Light-A-Video \cite{zhou2025light}  & \textbf{0.1313} & 0.1478 & 17.6803 & \underline{37.3366} & 542.1 & 5.5 & 2.94  \\
        I2VEdit \cite{ouyang2024i2vedit} & 0.1725 & 0.1490 & 22.4944 & 40.9389 &  821.1 & 8.3 & 1.82 \\
        DiffusionRenderer \cite{liang2025diffusionrenderer} & 0.1729 & \textbf{0.1621} & \textbf{9.7223} & 43.2499 &  \underline{460.2} & \underline{1.8} & \underline{3.27} \\
        Ours & \underline{0.1661} & \underline{0.1494} & \underline{15.2871} & \textbf{20.9532} & \textbf{456.1} & \textbf{0.9} & \textbf{4.08}  \\
        \bottomrule
    \end{tabular}

    \label{tab:quant_compare}
\end{table*}

\textbf{Evaluation Philosophy.}
It is important to clarify the goal of our experiments. We are not aiming to prove the superiority of our U-Net architecture in isolation. Instead, we evaluate our entire proposed pipeline, which includes a brief and efficient fine-tuning step on synthetic data as a core component. The central experimental question is whether this practical, data-efficient approach can outperform existing state-of-the-art methods when they are used in their standard, often zero-shot, capacity. This reflects a real-world use case and highlights the overall effectiveness of our method.

\textbf{Baselines.}
Our evaluation is designed to demonstrate ReLumix's versatility. Instead of comparing only to methods with identical inputs, we selected leading models from distinct relighting categories. This highlights how our framework can serve as an integration platform for:
\begin{itemize}
    \item \textbf{Text-to-Image Relighters:} IC-Light~\cite{iclight} and Light-A-Video~\cite{zhou2025light}, which use text prompts for artistic control.
    \item \textbf{Neural and Physically-Based Renderers:} DiffusionRenderer~\cite{liang2025diffusionrenderer}, which relights using environment maps (envmaps). Our framework can leverage the output of such renderers (or any physics-based method) as a reference frame.
    \item \textbf{General Video Editors:} I2VEdit~\cite{ouyang2024i2vedit}, another method that propagates edits from a reference frame. This provides a direct comparison for the core propagation task.
\end{itemize}
This diverse set of baselines shows that ReLumix can synergistically enhance text-, physics-, or image-driven workflows by extending their capabilities from static images to video.

\subsection{Synthetic Dataset CARLA Evaluation}
The CARLA dataset provides access to ground truth renderings, enabling a direct, objective evaluation of relighting fidelity. We use the CARLA test set, which comprises 18 unique scenes under 10 different lighting conditions. We provide the ground truth's first frame as the reference for our method and I2VEdit. Performance is measured using standard image fidelity metrics: SSIM, PSNR, and LPIPS.

As shown in Table~\ref{tab:CARLA_compare}, our method significantly outperforms the direct baseline I2VEdit across all metrics. This demonstrates the superiority of our specialized architecture for the task of illumination propagation. The qualitative results in Fig.~\ref{fig:qual_compare} (top row) visually corroborate these findings. More examples are available in the supplementary video.

\begin{figure}[!tb]
    \centering
    \includegraphics[width=\linewidth]{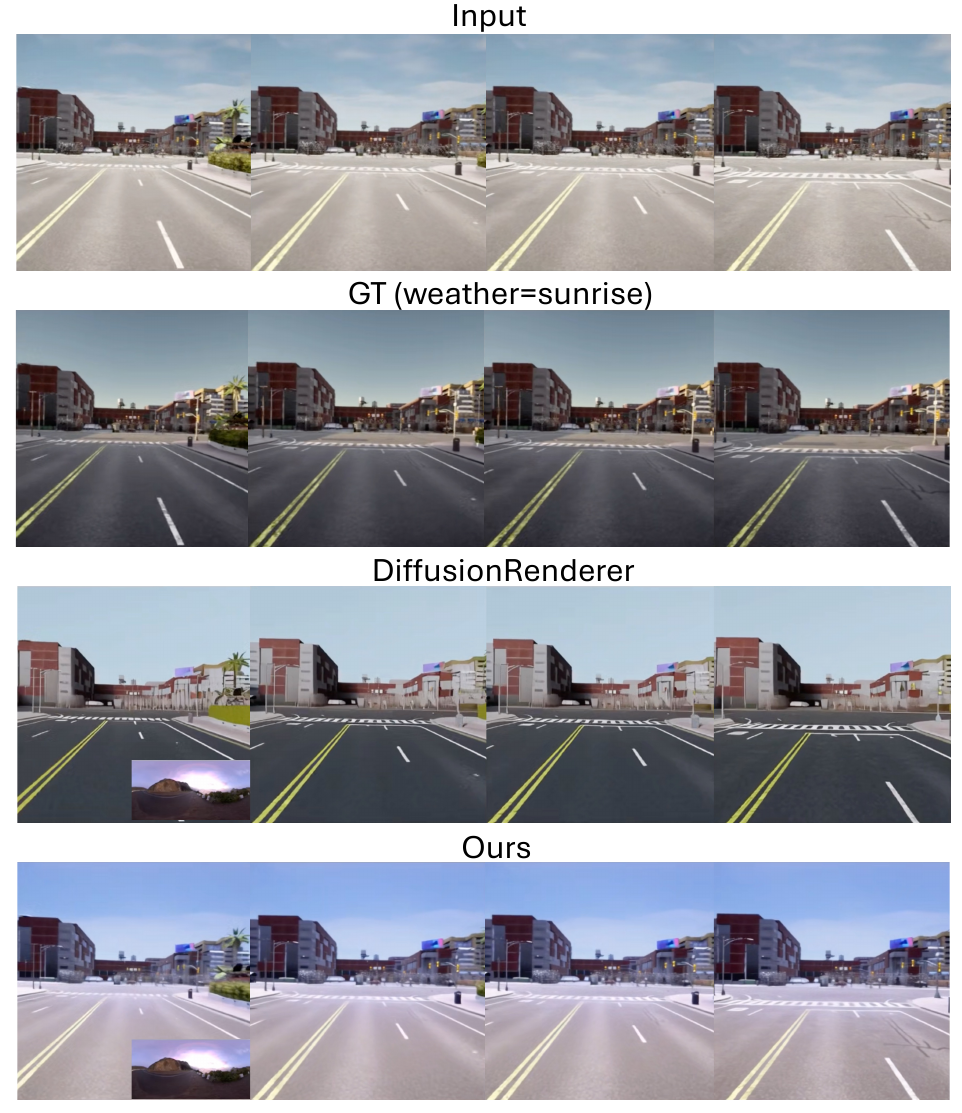}
    \caption{A qualitative comparison of our method and DiffusionRenderer using the same envmap. The bottom-right corner of the first image from the left shows the used envmap. As the CARLA simulator does not employ an envmap, the ground truth results are provided for reference only.}
    \label{fig:carla_envmap}
\end{figure}

\textbf{Integration with Physically-Based Rendering.}
A key strength of ReLumix is its ability to propagate not just stylized, but also physically plausible lighting. To demonstrate this, we conducted a targeted experiment against DiffusionRenderer, which uses an HDRI envmap for relighting. We first employed a physically-based inverse rendering method~\cite{wang2025materialist} to relight the first frame of a video using the \textit{exact same envmap}. This high-fidelity, physically-grounded image was then used as the reference for ReLumix. As shown in Fig.~\ref{fig:carla_envmap}, our method successfully propagates the complex, realistic lighting effects, preserving details and material interactions more faithfully than the stylized output of DiffusionRenderer. This highlights ReLumix's potential for use in production pipelines where physical accuracy is paramount.

\begin{figure}[tbh]
    \centering
    \includegraphics[width=\linewidth]{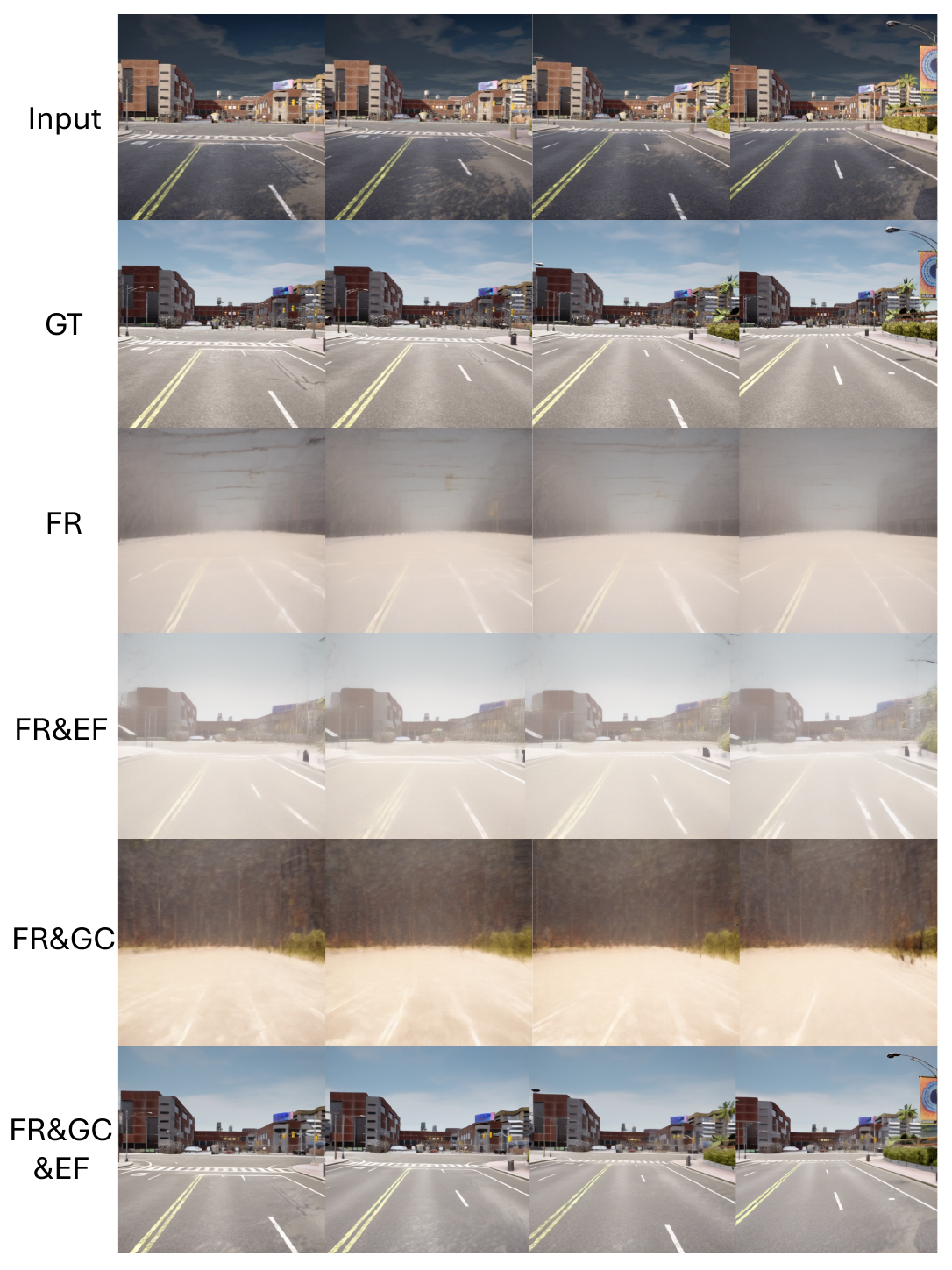}
    \caption{Ablation study qualitative results of different components. In the figure, \textit{FR} is Frame Replacement, \textit{EF} is Embedding Fusion, and \textit{GC} is Gated Cross-Attention.}
    \label{fig:ablation}
\end{figure}
\subsection{Real World Dataset DAVIS Evaluation}
To assess performance on in-the-wild videos, we used the DAVIS dataset. This evaluation focuses on temporal stability and perceptual quality, as ground truth is unavailable.

\textbf{Setup for a Creative Workflow.}
To simulate a practical artistic workflow, we used the popular and powerful IC-Light model to generate four distinct relighting styles for each video. The resulting image served as the reference frame for our method and I2VEdit. Light-A-Video was evaluated using the same prompts to ensure a fair comparison of artistic intent.

\textbf{Quantitative and Human Evaluation.}
We employed a suite of no-reference metrics to assess temporal consistency (Temporal LPIPS, Flow Consistency), color stability (Color Consistency), and overall perceptual quality (BRISQUE, CD-FVD). We also report the average per-frame processing time on an H100 GPU to quantify efficiency. As detailed in Table~\ref{tab:quant_compare}, our method achieves the best or second-best results across nearly all metrics, most notably being the fastest method while delivering top-tier quality (lowest BRISQUE and CD-FVD scores).

To complement the quantitative data, we conducted a user study with 18 participants who rated the overall satisfaction with the relighting results on a scale of 1 to 5. The results in Table~\ref{tab:quant_compare} show a strong preference for our method, confirming its superior visual fidelity and temporal stability. Qualitative comparisons are provided in Fig.~\ref{fig:qual_compare}, with extensive video results in our supplementary materials.

\begin{table*}[tbh]
    \centering
    \caption{Ablation study of different components. We report the p-values from statistical tests comparing the performance of model variants across three key metrics: SSIM, PSNR, and LPIPS. Our full model (FR \& GC \& EF) is compared against variants where one or both of the GC and EF modules are removed. Lower p-values indicate a statistically significant difference in performance. FR serves as the baseline model.}
    \label{tab:statistic_analysis}
    \setlength{\tabcolsep}{1mm}
    \begin{tabular}{c|ccc}
    \toprule
    \multirow{2}{*}{Comparison} & \multicolumn{3}{c}{Metric (p-value)} \\
    \cline{2-4}
     & SSIM & PSNR & LPIPS \\
    \midrule
    \textbf{FR \& GC \& EF} vs. FR (baseline) & $1.82 \times 10^{-12}$ & $5.46 \times 10^{-12}$ & $1.82 \times 10^{-12}$ \\
    \textbf{FR \& GC \& EF} vs. FR \& EF (-GC) & $9.09 \times 10^{-12}$ & $5.46 \times 10^{-12}$ & $1.82 \times 10^{-12}$ \\
    \textbf{FR \& GC \& EF} vs. FR \& GC (-EF) & $1.82 \times 10^{-12}$ & $1.82 \times 10^{-12}$ & $1.82 \times 10^{-12}$ \\
    \midrule
    FR (baseline) vs. FR \& EF & $0.413$ & $0.519$ & $0.034$ \\
    FR (baseline) vs. FR \& GC & $0.064$ & $1.46 \times 10^{-7}$ & $4.93 \times 10^{-6}$ \\
    FR \& EF vs. FR \& GC & $0.921$ & $1.04 \times 10^{-5}$ & $4.93 \times 10^{-6}$ \\
    \bottomrule
    \end{tabular}%
    % }
\end{table*}

\section{Ablation Study} \label{sec:ablation}
The quantitative evaluation of the ablation study for each component on CARLA testset is shown in Table~\ref{tab:CARLA_compare}, and the qualitative analysis is presented in Fig.~\ref{fig:ablation}. The results demonstrate that the desired results are only achievable through the full combination of FR, EF, and GC. 
Please refer to the supplementary video (Part 3) to see examples.

\textbf{Statistical analysis of different components. }
To quantitatively validate the contribution of our proposed FR, GC and EF modules, we conducted a rigorous statistical analysis, with the results summarized in Table~\ref{tab:statistic_analysis}. The analysis reveals that our full model, which integrates all components (FR \& GC \& EF), demonstrates a highly significant performance improvement over all ablated versions. Specifically, when compared with the baseline (FR), the variant without the GC module (FR \& EF), and the variant without the EF module (FR \& GC), the improvements across SSIM, PSNR, and LPIPS metrics are all statistically significant, with p-values consistently below $10^{-11}$. The inclusion of the GC module, in particular, yields substantial gains. The comparison between the baseline (FR) and the model with GC (FR \& GC) shows a highly significant difference for PSNR ($p < 1.46 \cdot 10^{-7}$) and LPIPS ($p < 4.93 \cdot 10^{-6}$), highlighting its crucial role in enhancing reconstruction quality and perceptual fidelity. Similarly, adding the EF module (FR vs. FR \& EF) also provides a statistically meaningful improvement in perceptual similarity, as indicated by the LPIPS p-value of $0.034$. These results statistically confirm that both the GC and EF modules are essential and complementary components, and their combination in our final model is critical for achieving superior performance.

\section{Limitations and Conclusion} \label{sec:limitations}
\begin{figure}
    \centering
    \includegraphics[width=\linewidth]{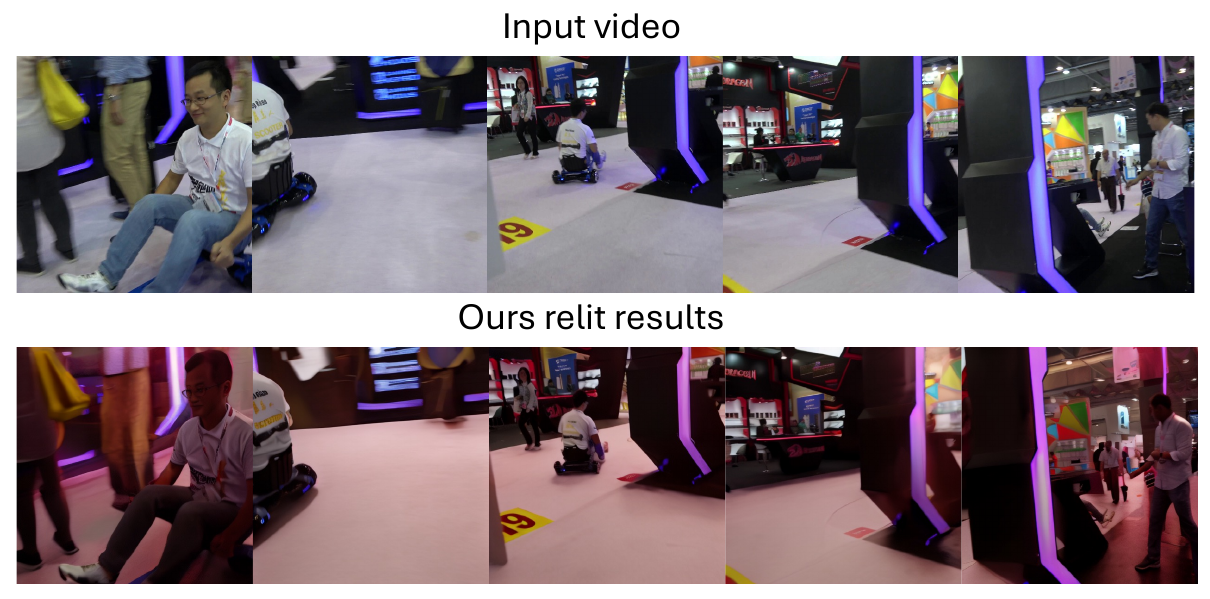}
    \caption{Limitation. Our method struggles to maintain lighting consistency during significant camera motion.}
    \label{fig:limit}
\end{figure}

Our work introduces ReLumix, a flexible and efficient framework that successfully bridges the gap between single-image relighting techniques and the demands of video post-production. By demonstrating robust sim-to-real generalization from a lightweight synthetic dataset, we present a practical path forward for data-efficient video editing. The modular design, which allows integration with both fast generative models and high-fidelity physically-based renderers, empowers artists with unprecedented flexibility. While our results are promising, we also recognize several limitations that open exciting avenues for future research.

\textbf{Limitations.}
The primary limitation of our current approach is its reliance on a single, static reference frame. As shown in Fig.~\ref{fig:limit}, this creates an information bottleneck when processing videos with significant camera motion or parallax, as new scene content is revealed that has no correspondence in the initial frame. While SVD's strong motion priors can handle moderate camera movement, large transformations can cause the propagated lighting to become less accurate or inconsistent over time. Similarly, our framework is not yet equipped to handle dynamic light sources within the scene, such as a moving spotlight or a passing vehicle's headlights.

In conclusion, ReLumix represents a important step towards practical and controllable video relighting. We believe that hybrid approaches like ours, which combine the power of large-scale generative models with the precision of classic graphics principles and artist-in-the-loop workflows, will continue to redefine the landscape of digital content creation.

%% file: sec/X_Appendix.tex
\twocolumn[{
\begin{center}
\Large\textbf{Supplementary Materials}
\vspace{2 ex}
\end{center}
}]

\section{Integration of Physics-Based Inverse Rendering for Video Relighting}
\label{sec:physically_based_integration}
The proposed ReLumix framework incorporates physics-based inverse rendering techniques to achieve high accurate relighting of the initial frame in a video sequence. This approach ~\cite{wang2025materialist} involves estimating material properties and geometric configurations through iterative optimization. While this method is computationally intensive, its physical plausibility makes it ideal for generating accurate lighting conditions for subsequent video propagation.

The inverse rendering process begins with material property estimation. Given an input image $I_0$, we first predict scene properties using a pretrained network $\mathcal{P}$:
\begin{equation}
    A_p, R_p, M_p, N_p, D_p = \mathcal{P}(I_0),
\end{equation}

where $A_p$, $R_p$, and $M_p$ represent the albedo, roughness, and metallic maps, while $N_p$ and $D_p$ denote the surface normal and depth maps. These predictions are refined through optimization that minimizes the reconstruction error between the rendered and original images:
\begin{equation}
    \min_{A_p, R_p, M_p, N_p, D_p} \mathcal{L}_p(I_0, I_p),
\end{equation}

where $I_p$ is the rendered frame using the predicted properties, and $\mathcal{L}_p$ is defined as:
\begin{equation}
    \mathcal{L}_p = \mathcal{L}_{\text{re}} + \delta \mathcal{L}_{\text{cons}},
\end{equation}

with $\mathcal{L}_{\text{re}}$ as the reconstruction loss and $\mathcal{L}_{\text{cons}}$ enforcing consistency between optimized and predicted material properties.

For environment map optimization, we employ a position-embedded MLP to model the incident radiance $E(\omega_i)$:
\begin{equation}
    E(\omega_i) = \text{MLP}_{\text{env}}(\gamma(\mathbf{e}); \theta_{\text{env}}),
\end{equation}

where $\gamma(\cdot)$ is a positional encoding function. The rendering equation for reflected radiance becomes:
\begin{equation}
    L_o(\mathbf{x}, \vec{\omega}_o) = \int_{\Omega} f_r(\mathbf{x}, \vec{\omega}_i, \vec{\omega}_o) L_i(\mathbf{x}, \vec{\omega}_i) (\vec{\omega}_i \cdot \vec{n}) \, d\omega_i,
\end{equation}

where $f_r$ is the BRDF, $L_i$ is the incident radiance, and $\vec{n}$ is the surface normal.

Material property optimization further refines the estimates by minimizing:
\begin{equation}
    \min_{A, R, M} \mathcal{L}_{\text{Mat}}(L_o, I),
\end{equation}

with $\mathcal{L}_{\text{Mat}} = \mathcal{L}_{\text{re}} + \delta \mathcal{L}_{\text{cons}}$, where $\mathcal{L}_{\text{cons}}$ ensures alignment with initial predictions. The final material maps $A^*, R^*, M^*$ are then used to compute the relit frame $I_0^*$ via:
\begin{equation}
    I_0^* = \int_{\Omega} \mathcal{M}(\omega_i) \lambda(\omega_i) (\omega_i \cdot N^*) \, d\omega_i,
\end{equation}

where $\lambda(\omega_i) = E(\omega_i)$ and $\mathcal{M}(\omega_i)$ represents the material response using the Disney BRDF~\cite{burley2012physically}.

By integrating this physics-based inverse rendering pipeline, ReLumix achieves accurate initial frame relighting while maintaining computational feasibility through its modular design. The optimized material properties and environment map from this stage serve as the foundation for subsequent video propagation, ensuring temporal consistency and physical accuracy across the entire sequence.

\section{More Experiments Results}

\textbf{Standard Deviation of Quantitative Evaluation over \textit{CARLA Relight}}

We provide standard deviations alongside the mean values to quantify the consistency on \textit{CARLA Relight} in Table \ref{tab:quant_std}.

\begin{table}[!h]
    \centering
    \caption{Standard Deviation of quantitative evaluation study over \textit{CARLA Relight}.}
    \begin{tabular}{c|ccc}
    \toprule
     Methods& SSIM & PSNR & LPIPS \\ 
     \midrule
     DiffusionRenderer & 0.072 & 1.451 & 0.024 \\
     \hline
     I2VEdit & 0.200 & 5.953 & 0.170 \\
    \hline
    Ours (w/ FR) & 0.227 & 3.656 & 0.086 \\
    Ours (w/ FR\&EF) & 0.524 & 3.791 &0.091 \\
    Ours (w/ FR\&GC) & 0.257 & 3.791 & 0.128\\
    Ours (w/ FR\&GC\&EF) & 0.125 & 2.554 & 0.057 \\
    \bottomrule
    \end{tabular}
    \label{tab:quant_std}
\end{table}

\textbf{Standard Deviation of Quantitative Evaluation}
In addition to the average values presented in the paper, we also provide the standard deviation for reference as shown in Table \ref{tab:davis_std}. As can be seen, the standard deviation of our method falls within a reasonable range.

\begin{table*}[!h]
    \centering
    \caption{Standard Deviation of Quantitative Evaluation Results on real world videos (DAVIS).}
    % \resizebox{\linewidth}{!}{
    \setlength{\tabcolsep}{1mm}
    \begin{tabular}{c|cccc}
        \toprule
        \multirow{3}{*}{Method}&\multicolumn{4}{c}{Standard Deviation of Quantitative Evaluation}  \\
        \cline{2-5}
        & \makecell{Temporal \\ LPIPS} & \makecell{Flow \\ Consistency} & \makecell{Color \\ Consistency} & \makecell{Brisque} \\
        \midrule
        Input (Reference) & 0.107 & 0.080 & 5.070 & 10.60  \\
        \hline
        IC-Light  & 0.111 & 0.041 & 7.020 & 11.71  \\
        Light-A-Video  & 0.080 & 0.087 & 7.180 & 11.93    \\
        I2VEdit & 0.057 & 0.105 & 7.424 & 12.17  \\
        DiffusionRenderer & 0.105 & 0.092 & 4.758 & 14.69  \\
        Ours & 0.103 &0.091 & 5.746 & 12.46    \\
        \bottomrule
    \end{tabular} %}
    \label{tab:davis_std}
\end{table*}

\section{Prompt for Text-based Video Relighting}
\label{sec:prompt_engineering}
To facilitate the evaluation of video relighting capabilities, we design a set of structured prompts that incorporate diverse lighting conditions and spatial configurations. These prompts are specifically tailored for the IC-Light \cite{iclight} and Light-A-Video \cite{zhou2025light} frameworks, ensuring consistency in experimental settings. The prompt formulation involves two key components: negative prompts to suppress undesirable artifacts and positive prompts to specify desired lighting attributes . 

The negative prompt is defined as \textit{"bad quality, worse quality"} to mitigate the generation of low-resolution or distorted outputs. For the positive relighting prompts, we introduce a series of lighting scenarios that capture distinct atmospheric and directional characteristics. These include: (1) a \textit{"warm atmosphere, neon light, city"} configuration to simulate urban environments with artificial illumination; (2) a \textit{"cool blue lighting, cyberpunk style"} prompt for futuristic, high-contrast scenes; (3) a \textit{"home atmosphere, cozy illumination"} scenario to replicate domestic lighting conditions; and (4) a \textit{"sunset over the sea, warm light, soft light"} configuration to model diffuse, ambient lighting. 

The spatial lighting configuration is further refined through the \texttt{bg\_source} parameter, which determines the directional origin of the illumination. This parameter is set to \textit{"Right Light"} to ensure consistent directional bias across all experiments. Additional settings include a fixed random seed (42) to enable reproducibility, a resolution of $512\times512$ pixels, and a diffusion process with 25 denoising steps. The text guidance scale (2.0) and noise strength (0.9) are calibrated to balance prompt adherence and temporal stability. Finally, the we skip the background removal to retain background elements during the relighting process, ensuring spatial coherence in generated sequences
\section{Evaluation Metrics}
\label{sec:metrics}

To quantitatively assess the quality of the video relighting results, especially in the absence of ground truth data, we employ a set of no-reference metrics. These metrics are designed to evaluate various aspects of video quality, including temporal stability, motion coherence, color fidelity, and overall perceptual quality.

\textbf{Temporal LPIPS}
Learned Perceptual Image Patch Similarity (LPIPS) is utilized to measure the perceptual dissimilarity between consecutive frames. This metric is based on a deep neural network (AlexNet) trained to approximate human perceptual judgments. We calculate the LPIPS score between each pair of adjacent frames, $I_t$ and $I_{t+1}$, throughout the video sequence. The final Temporal LPIPS score is the average of these individual scores. A lower value indicates greater perceptual similarity between frames, signifying higher temporal stability and fewer flickering artifacts.

\textbf{Flow Consistency}
To evaluate the coherence of motion, we compute a flow consistency score based on optical flow. We first detect and track salient feature points between consecutive frames using the Lucas-Kanade algorithm \cite{lucas1981iterative}. For the resulting flow vectors, we calculate their magnitudes and angles. The consistency is then quantified by computing the standard deviation of these magnitudes ($\sigma_{mag}$) and angles ($\sigma_{angle}$) over the entire video. The final score is defined as $(1 + \sigma_{mag} + \sigma_{angle})^{-1}$. A higher score indicates more stable and consistent motion patterns, suggesting that the relighting process has not introduced unnatural motion jitter.

\textbf{Color Consistency}
Color consistency is assessed by analyzing the stability of chromatic information across frames. Each frame is converted to the CIELAB color space, which separates luminance (L*) from chrominance (a*, b*). We then compute the magnitude of the chromaticity components, $\sqrt{(a^*)^2 + (b^*)^2}$, for each pixel. The standard deviation of these chromaticity magnitudes is calculated for each frame, and the final Color Consistency score is the average of these standard deviations over the sequence. A lower score signifies less fluctuation in the color distribution, indicating that the relighting has preserved a consistent color palette.

\textbf{BRISQUE}
The Blind/Referenceless Image Spatial Quality Evaluator (BRISQUE) \cite{mittal2012no} is a no-reference image quality assessment model that evaluates the overall perceptual quality of individual frames. It operates by extracting natural scene statistics from each frame and using them to predict a quality score. This metric effectively captures a wide range of distortions, including noise, blur, and compression artifacts, without requiring a pristine reference image. We compute the BRISQUE score for every frame and report the average value. A lower BRISQUE score corresponds to a higher perceptual quality of the video.

\section{Details of Human Evaluation on Real World Dataset}
A user study was conducted to evaluate the performance of our method against several state-of-the-art techniques. 18 volunteers participated in the evaluation. The participant pool was diverse, comprising six individuals from non-engineering backgrounds and twelve from engineering disciplines. Among the engineering-focused participants, four possessed prior expertise in the video relighting task.

For the evaluation, participants were presented with video relighting results generated by five different methods. The specific method corresponding to each video was not disclosed to the participants to prevent bias. The videos were displayed simultaneously in a side-by-side arrangement, synchronized frame-by-frame with the original video. This setup allowed for direct and immediate comparison of the relighting effects. Participants were given full control over the playback, including the ability to play, pause, and seek forward or backward through the videos. No time constraints were imposed, and participants were encouraged to take as much time as needed to form their judgments. They were also permitted to assign identical scores to methods they perceived as having comparable performance.
A specific instruction was provided regarding the evaluation of one of the methods, DiffusionRenderer. Due to its use of environment maps for relighting, its visual style differs significantly from the other techniques. Participants were explicitly asked to disregard these stylistic differences and focus their assessment solely on the quality and fidelity of the relighting itself. 

\section{Implementation Details}

The proposed framework is constructed upon a diffusion-based architecture, with a structured design that ensures temporal coherence and high-quality video relighting. The core components of the model are as follows:  

The backbone network employs a \texttt{VideoUNet} as the primary architecture, specifically tailored for video data processing while preserving temporal consistency. This network features an input channel configuration of 12, derived from the concatenation of video frames and conditioning inputs. The output channels are set to 4, corresponding to the predicted noise or video frames. The model channels are initialized at 320, with channel multipliers of $[1, 2, 4, 4]$, enabling hierarchical feature extraction. Attention mechanisms are integrated at resolutions of $[4, 2, 1]$, and a gated cross-attention mechanism is incorporated to enhance temporal stability during video generation.  

The conditioning module utilizes a \texttt{GeneralConditioner} to process multiple input modalities, including reference frames, timestep embeddings, and auxiliary information. Each conditioning input is processed through a frozen encoder, such as \texttt{FrozenOpenCLIPImageEmbedder}, to extract discriminative features. This ensures the model can effectively incorporate contextual information during the relighting process.  

The denoising process is implemented with a \texttt{VScalingWithEDMcNoise} configuration, which dynamically adjusts noise levels during both training and inference phases. This adaptive scaling improves the robustness of the diffusion process, particularly for complex lighting scenarios.  

For inference, the \texttt{EulerEDMSampler} is employed with 25 denoising steps, balancing computational efficiency and output quality. This sampler ensures the generation of temporally coherent video sequences while maintaining high fidelity.  

\textbf{Training Configuration}  
The training process is configured with an Adam optimizer, utilizing a base learning rate of $2 \times 10^{-5}$. The loss function employs \texttt{StandardDiffusionLoss} with \texttt{EDMWeighting}, which prioritizes the top $10\%$ of losses during the initial 5000 training steps to enhance stability. The batch size is set to 2 videos per batch, with each video containing 14 frames. The model is trained for up to 10,000 steps, allowing sufficient convergence for complex relighting tasks.  

\textbf{Dataset and Preprocessing}  
The CARLA dataset is utilized for training, with a preprocessing pipeline designed to ensure compatibility and consistency. We divided the dataset into train, validation, and test sets in the ratio of 0.7, 0.2, and 0.1, respectively. All videos are resized to $512 \times 512$ resolution to align with the model's input requirements. The first frame of each video is replaced with a relit reference frame to enforce photometric alignment. To increase dataset diversity, random temporal reversals are applied with a probability of 0.2. 

\textbf{Inference Settings}  
During inference, the \texttt{EulerEDMSampler} is used with 25 denoising steps to balance efficiency and quality. Noise levels are dynamically scaled using the \texttt{EDMDiscretization} method, with a maximum sigma value of 700.0. A linear prediction guider is employed to control the scale of generated outputs, with a range of $[1.0, 2.5]$. These settings ensure the model produces temporally coherent and visually realistic relit videos while adhering to the constraints of real-world deployment.